\documentclass{INTERSPEECH2023}

\interspeechcameraready 

\usepackage{cite}

\title{QuickVC: Any-To-Many Voice Conversion Using Inverse Short-Time Fourier Transform for Faster Conversion}
\name{Houjian Guo$^1$$^2$, Chaoran Liu$^3$, Carlos Toshinori Ishi$^1$$^3$, Hiroshi Ishiguro$^2$$^3$}
\address{
  $^1$Interactive Robot Research Team, Guardian Robot Project, RIKEN, Japan\\
  $^2$Graduate School of Engineering Science, Osaka University, Japan\\
  $^3$Advanced Telecommunications Research Institute International, Japan}
\email{}

\begin{document}

\maketitle
 
\begin{abstract}
With the development of automatic speech recognition (ASR) and text-to-speech (TTS) technology, high-quality voice conversion (VC) can be achieved by extracting source content information and target speaker information to reconstruct waveforms. However, current methods still require improvement in terms of inference speed. In this study, we propose a lightweight VITS-based VC model that uses the HuBERT-Soft model to extract content information features without speaker information. Through subjective and objective experiments on synthesized speech, the proposed model demonstrates competitive results in terms of naturalness and similarity. Importantly, unlike the original VITS model, we use the inverse short-time Fourier transform (iSTFT) to replace the most computationally expensive part. Experimental results show that our model can generate samples at over 5000 KHz on the 3090 GPU and over 250 KHz on the i9-10900K CPU, achieving competitive speed for the same hardware configuration.\footnote{The audio samples are available at \url{https://quickvc.github.io/quickvc-demo}}
\end{abstract}

\begin{figure*}[htbp]
    \begin{minipage}[t]{0.6\linewidth}
        \centering
        \includegraphics[width=\textwidth]{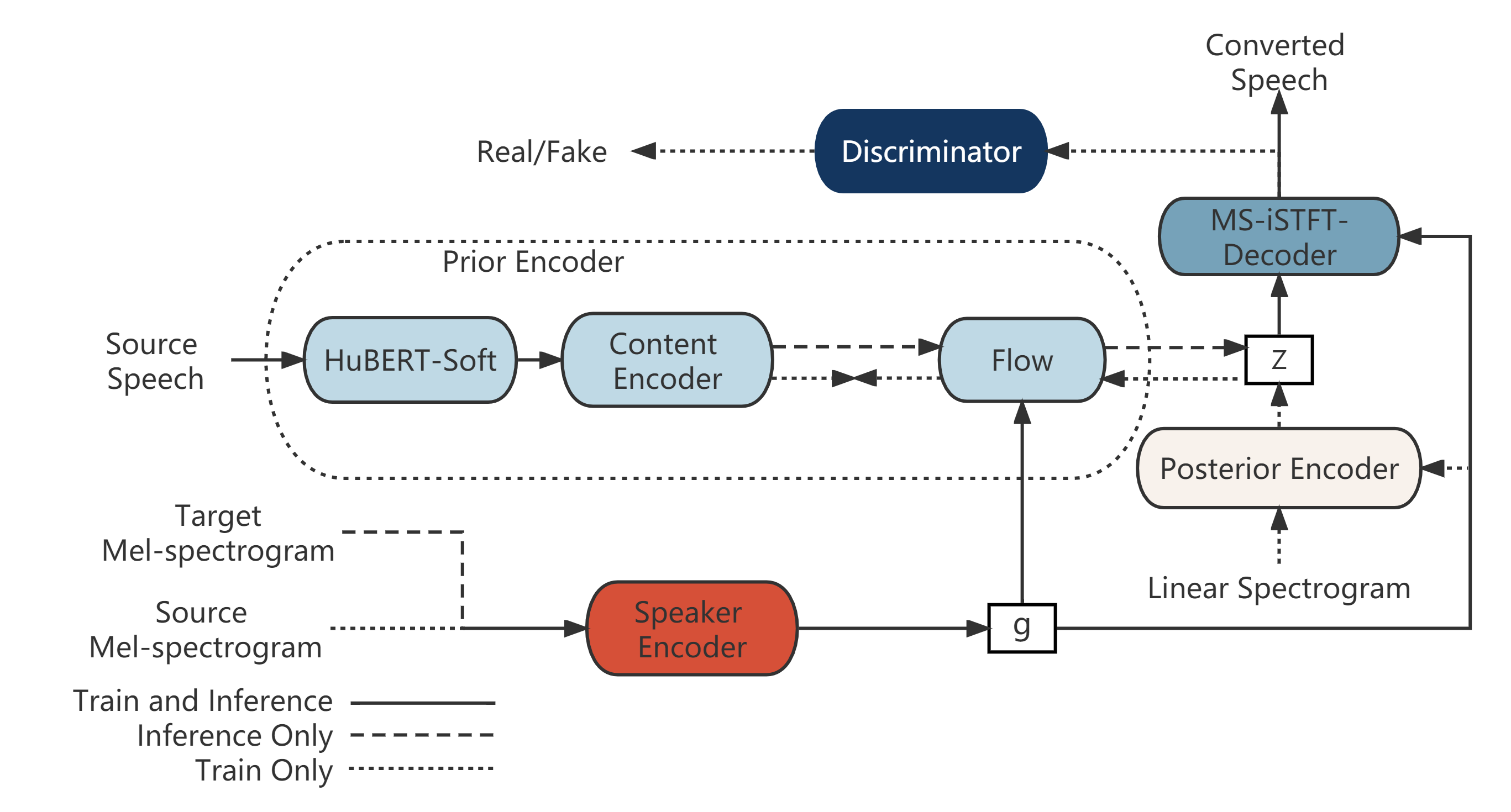}
        \centerline{(a) QuickVC model}
    \end{minipage}%
    \begin{minipage}[t]{0.4\linewidth}
        \centering
        \includegraphics[width=\textwidth]{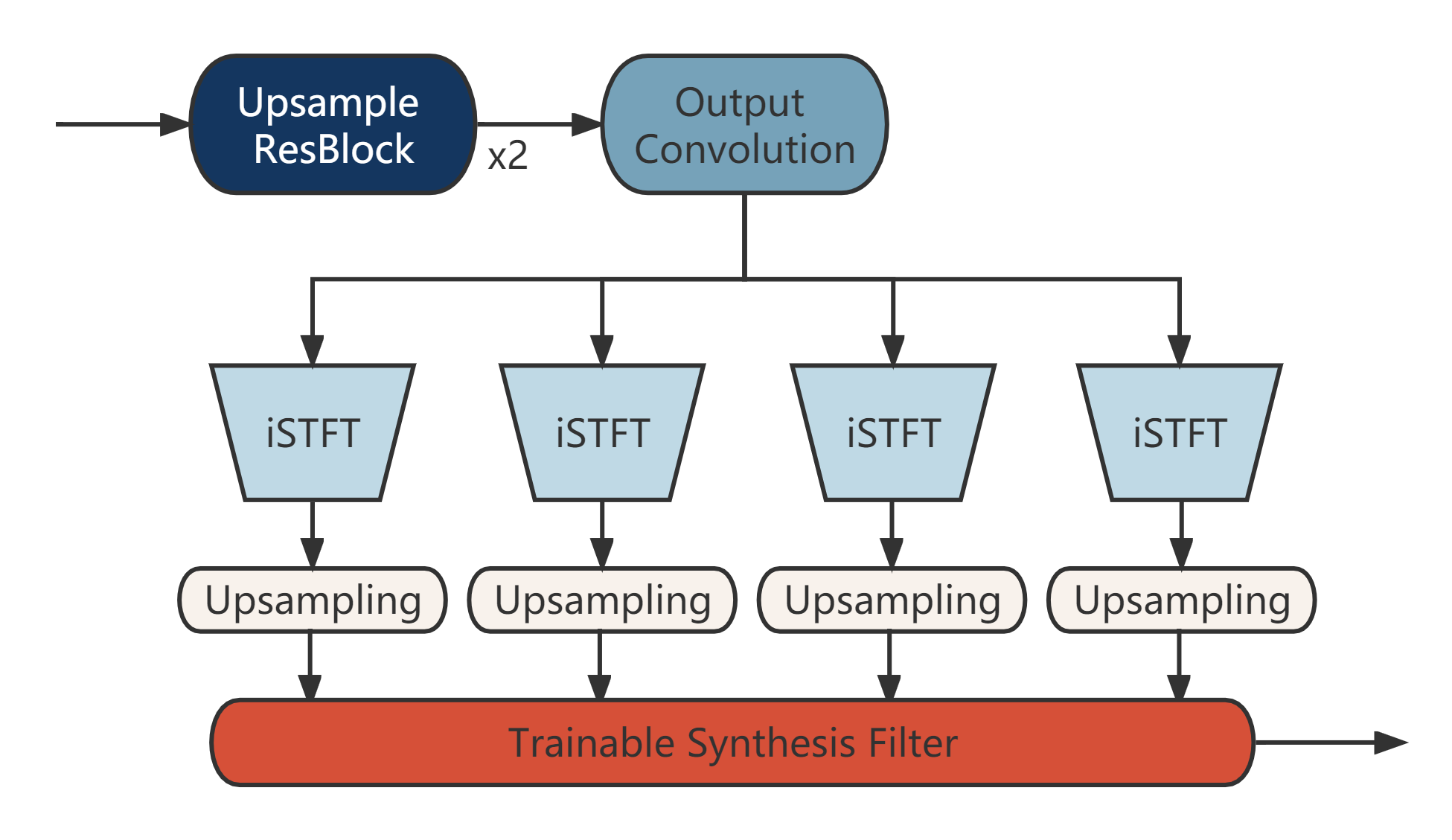}
        \centerline{(b) MS-iSTFT-Decoder}
    \end{minipage}
    \caption{Overview network architecture of QuickVC. (a) Here $g$ denotes speaker embedding, and $z$ denotes the latent variable. (b) In the MS-iSTFT-Decoder, synthesized waveforms are integrated by a trainable synthesis filter. $\times n$ means the indicated block repeats n times. }
\end{figure*}

\section{Introduction}
Voice conversion (VC) is a technique that modifies a speaker's voice to sound like that of another speaker while preserving the original meaning and content of the speech \cite{kaneko2018cyclegan}. In this article, we aim at any-to-many voice conversion, which is the conversion of an arbitrary speaker's source speech signal to a target speaker appearing in the training set.

A typical approach of any-to-many VC is feature disentangling. The content feature information and the speaker feature information in the speech are extracted separately and used to reconstruct the speech \cite{qian2019autovc}. The results of this method depend on whether the obtained content features do not contain the original speaker feature information, while not losing information about the speech content.


Techniques from the ASR domain are widely used to extract linguistic content from speech while ignoring speaker-specific details. VC based on phonetic posteriorgrams (PPGs) draws much attention \cite{liu2021any,li2021ppg}. PPGs are intermediate results from a speaker-independent ASR system, representing the posterior probability of phonetic classes at the frame level \cite{hazen2009query}. PPGs are independent of speaker and language, making them suitable for VC. However, the accuracy of the PPG-based VC model largely depends on the precision of the ASR model used to extract the PPGs. In addition to ASR, transfer learning from TTS methods has been employed to obtain linguistic representations for VC \cite{zhang2021transfer}. However, these methods require a large amount of annotated data containing text to train the ASR or TTS model. There are also studies that do not require text annotated data, such as Cyclegan-VC \cite{kaneko2018cyclegan} and StarGAN \cite{kaneko2019stargan}, which are GAN-based models, and AutoVC \cite{qian2019autovc}, which is an AutoEncoder. But the speech they generate is relatively poor in terms of quality \cite{zhao2020voice}.

Recently there has been a lot of new research on VC that attempts to obtain feature vectors of speech by means of self-supervised learning (SSL) models \cite{huang2021any,van2022comparison,lifreevc}. By obtaining the content representation of speech from these feature vectors, speech can be reconstructed to achieve VC or singing voice conversion (SVC)\footnote{https://github.com/innnky/so-vits-svc/tree/32k}. These studies achieve very good results in terms of quality and are close to the TTS models.

The development of TTS models has helped to produce high-quality speech based on content features. TTS models such as Tacotron2 \cite{tacotron2} and Fastspeech \cite{ren2019fastspeech} have the ability to synthesize naturalistic speech. They have been applied in the field of VC \cite{zhao2021towards, van2022comparison}. However, these TTS methods are two-stage, synthesizing acoustic features first and then using a vocoder to synthesize waveforms from the predicted acoustic features. VITS is an end-to-end TTS model that enhances the quality of the reconstructed waveform through adversarial training \cite{kim2021conditional}. By applying VITS to VC, separate training of the VC model and the vocoder can be avoided.

Although there are VC models that are capable of producing speech of good perceptual quality, there is still a lack of research on high-quality and fast VC. In practice, if you want to achieve real-time VC, you need to process the input speech frame by frame \cite{saeki2020real}, so the inference speed of the models is important for real-time VC. The faster the model performs this, the less delay will be experienced when converting speech. However, previous research in real-time VC has not been good enough in terms of the naturalness and similarity of the synthesized speech \cite{saeki2021real}. Now that high-quality VC is possible, there is a need for VC models that can maintain high quality while allowing fast inference.

In this study, we implement a fast and high-quality voice conversion model. The main contributions of this paper can be summarized as follows

\begin{itemize}
\item We propose the QuickVC model, which combines the high-quality speech synthesis model VITS with the speech content feature extraction model HuBERT-Soft to achieve high-quality any-to-many speech conversion.
\item In the VITS-based speech reconstruction part, the decoder structure is lightened to speed up the model. The inference speed of the model on the CPU is up to 280KHz.
\item To make the model pay more attention to the content information in the input features, a data augmentation method is used during the training process, which improves the naturalness and similarity of the results.
 \end{itemize}

\section{Method}
\subsection{Motivation and strategy}

QuickVC is inspired by VITS \cite{kim2021conditional}, Soft-VC \cite{van2022comparison} and MS-iSTFT-VITS \cite{kawamura2022lightweight} respectively. The backbone of QuickVC is inherited from VITS, which adopts variational inference, augmented with normalizing flows and an adversarial training process. We chose VITS as the basis for our VC system because of its ability to produce excellent speech synthesis and its non-autoregressive design, which is optimal for fast inference. However, unlike VITS, QuickVC's prior encoder takes the raw waveforms as input rather than the phonemes. The prior encoder refers to HuBERT-Soft in Soft-VC to obtain speaker-independent content feature information. Speaker embeddings are extracted by a speaker encoder to perform multi-speaker VC. In addition, the decoders of QuickVC and VITS are different. To achieve faster inference, we refer to the MS-iSTFT-VITS, which is a single-speaker TTS model that replaces the original Hifigan vocoder-based decoder \cite{kong2020hifi} with multi-stream iSTFTNet \cite{kaneko2022istftnet}.

\subsection{Model architecture}

QuickVC model consists of a speaker encoder, a prior encoder, a posterior encoder, an MS-iSTFT-Decoder, and a discriminator, where the architecture of the posterior encoder and discriminator follow VITS. We will focus on describing the prior encoder, speaker encoder, and  MS-iSTFT-Decoder in the following subsections.
\subsubsection{Prior encoder}

The prior encoder consists of HuBERT-Soft, a content encoder, and a  normalizing flow. As the input is no longer text but speech, the text encoder in VITS becomes HuBERT-Soft and the content encoder. HuBERT-Soft is a kind of feature extractor using HuBERT-Base as a backbone \cite{hsu2021hubert}. HuBERT-Soft takes the raw waveform as input and produces a 256-dimensional content feature.  The content encoder is based on the posterior encoder in VITS. The content encoder can produce the mean and variance of the normal posterior distribution.  The normalizing flow follows VITS and is a stack of affine coupling layers. The flow is used to improve the complexity of the prior distribution, conditioned on the speaker embedding $g$, to achieve any-to-many VC.

\subsubsection{Speaker encoder}
The speaker encoder generates an encoded speaker representation from an utterance. It is trained from scratch together with the rest of the model. The network structure of the speaker encoder contains one layer of LSTM structure and one layer of fully connected layers, which is referenced from  \cite{liu2021any}. We extract mel-spectrograms from the audio signal and use them as input to the speaker encoder. We assume that the output of the content encoder does not contain any speaker information. The model then replaces the missing speaker information based on the input from the speaker encoder to synthesize speech.
\subsubsection{MS-iSTFT-Decoder}
The decoder module is considered to be the biggest bottleneck in VITS according to the previous study  \cite{kawamura2022lightweight}. The decoder architecture in VITS is based on the HiFi-GAN vocoder, which uses a repeated convolution-based network to upsample the input acoustic features. We refer to the decoder architecture in MS-iSTFT-VITS, the decoder performs the following steps in sequence. First, the VAE latent variable $z$ is conditioned on the speaker embedding $g$, then upsampled by a sequence of upsample convolutional residual blocks (ResBlock)  \cite{he2016deep}. The upsampled $z$ is then projected to the magnitude and phase variables for each sub-band signal. Then, using the magnitude and phase variables, the iSTFT operation is performed to generate each sub-band signal. Finally, these sub-band signals are upsampled by inserting zeros between samples to match the sampling rate of the original signal and then integrated into full-band waveforms using a trainable synthesis filter.
\subsection{Training strategy}
\subsubsection{Data augmentation}
In the original Soft-VC, HuBERT-Soft was applied to the any-to-one VC task. To investigate whether the self-supervised features obtained by HuBERT-Soft still contain the features of the original speaker, we performed spectrogram-resize (SR) based data augmentation on the speech in the training dataset using the method in  \cite{lifreevc}. After data augmentation, we obtained speech with the same content and speed, but with a different speaker tone. We hope that by training with augmented speech, the content encoder in the model will learn better to extract the unchanged content information. 
\subsubsection{Speaker encoder input}
In the training process, the speaker encoder is first fed with a different utterance from the same target speaker. This helps the speaker encoder in the model to follow the speaker-related information in the speech at the start of training. In the final steps of training, the target speech input is used to fine-tune the speaker encoder so that the model output is more similar to the reference speech.

\subsubsection{Training loss}

Following VITS, the QuickVC model combines VAE and adversarial training in the training process. For the generator part, the loss can be expressed as:
\begin{equation}
  {\cal L}_{v a e}={\cal L}_{r e c o n}+{\cal L}_{k l}+{\cal L}_{a d v}(G)+{\cal L}_{f m}(G),
  \label{eq0}
\end{equation}
where the ${\cal L}_{r e c o n}$ is reconstruction loss, which is the L1 distance between target mel-spectrogram $x_{m e l}$ and predicted mel-spectrogram ${\hat{x}}_{m e l}$:
\begin{equation}
  L_{r e c o n}=\Vert x_{m e l}-{\hat{x}}_{m e l}\Vert_{1} .
  \label{eq1}
\end{equation}
The ${\cal L}_{k l}$ is KL loss, which is the KL divergence between the prior distribution $p_{\theta}(z|c)$ and posterior distribution $ q_{\phi}(z|x_{l i n})$, where
\begin{equation}
  {{q_{\phi}(z|x_{l i n})=N(z;\mu_{\phi}(x_{l i n}),\sigma_{\phi}(x_{l i n})),}} 
  \label{eq2}
\end{equation}
\begin{equation}
  p_{\theta}(z|c)=N(f_{\theta}(z);\mu_{\theta}(c),\sigma_{\theta}(c))|d e t\frac{\partial f_{\theta}(z)}{\partial z}|.
  \label{eq3}
\end{equation}
The $x_{l i n}$ is the linear spectrogram of the source speech.  $f_{\theta}$ means the normalizing flow. The condition $c$ is the content information extracted from the source waveform by HuBERT-Soft.

The adversarial loss ${\cal L}_{a d v}(G)$ and the feature matching loss ${\cal L}_{f m}(G)$ is the adversarial training loss \cite{mao2017least}, which is same as the VITS.
To adopt adversarial training in our learning system, we add a discriminator that discriminates between the output generated by the decoder and the ground truth waveform, the loss for the discriminator is ${\cal L}_{a d v}(D)$, which is also same as the VITS.

\section{Experiment}

\subsection{Dataset}
We perform experiments on VCTK  \cite{veaux2017cstr}, LibriSpeech  \cite{panayotov2015LibriSpeech}, and LJ Speech \cite{ito2017lj}. Only the VCTK corpus is used for training, which contains 44 hours of utterances from 107 English speakers with different accents. We use the LibriSpeech and the LJ Speech dataset as source speech to test the ability of the model in the any-to-many VC scenario.
\subsection{Experimental setups}

For our experiments, a sampling rate of 16,000 Hz is used. We randomly split the utterances of a speaker into training and test sets in a 9:1 ratio. Both linear spectrograms and 80-band mel-spectrograms are computed using short-time Fourier transform, with FFT, window, and hop size set to 1280, 1280, and 320, respectively. Our models are trained on a single NVIDIA 3090 GPU for up to 600k steps, with a batch size of 64 and a maximum segment length of 512 frames. The MS-iSTFT-Decoder uses the same FFT size, hop length, window length, number of sub-bands, and kernel size as those in MS-iSTFT-VITS. The upsampling scale of the two residual blocks is [5,4].  With the same weight decay and learning rate as VITS, the AdamW optimizer  \cite{loshchilovdecoupled} is used.


\subsection{Baselines}
We select three baseline VC models to compare with our proposed model.

\begin{itemize}
\item Diff-VCTK: a high-quality VC model that uses the diffusion model to generate high-quality speech \cite{popov2021diffusion}.
\item BNE-PPG-VC: a model that uses PPGs to obtain speech content features that do not include speaker features \cite{liu2021any}.
\item VQMIVC: a model that uses vector quantization for content encoding and introduces mutual information as a correlation metric \cite{wang2021vqmivc}.
\end{itemize}

In addition to the baseline model, our experiments also compare two versions of QuickVC.
\begin{itemize}
\item QuickVC-nosr does not employ any data augmentation.
\item QuickVC-sr uses SR-based data augmentation when preparing the training dataset.
\end{itemize}

\subsection{Evaluation metrics}

To obtain a comprehensive evaluation of the model, subjective and objective experiments were conducted separately. 

For the subjective experiments, following Wester et al.  \cite{wester2016analysis}, we adopted Mean Opinion Score (MOS) to compute the naturalness and similarity scores of the converted utterances as subjective metrics. We invited 80 subjects on Amazon Mechanical Turk to participate in all the experiments. Forty subjects individually evaluated the naturalness of 6 original utterances from the dataset and 40 converted utterances. Forty subjects individually evaluated the similarity of 2 utterances from each of the 6 speakers in the dataset and the similarity of the 30 converted utterances to the utterances of the target speakers. 

For the objective experiments, we test speaker similarity,  intelligibility and inference speed. We used the trained model Resemblyzer\footnote{https://github.com/resemble-ai/Resemblyzer} to score the speaker similarity between the converted speech and the target speech. We randomly select 400 utterances (200 from VCTK, 200 from LibriSpeech) as source speech and 6 speakers from VCTK as target speakers.  Intelligibility is scored by the word error rate (WER) and character error rate (CER) between the source speech and the converted speech \cite{DBLP:conf/icmi/ZhaoLZYYDH21}. WER and CER were obtained using the ASR model HuBERT-Large\footnote{https://huggingface.co/facebook/hubert-large-ls960-ft}. The ASR model is trained with LibriSpeech, so only the utterances converted from LibriSpeech are used to score  intelligibility. The inference speed of the models was measured separately for Intel(R) Core(TM) i9-10900K CPU @ 3.70GHz and NVIDIA GeForce RTX 3090 GPU. The measurement of VC time refers to the time taken from the model's input of the source voice to the model's output of the converted voice.

\subsection{Results}

\subsubsection{Naturalness and similarity of speech}

The experimental results of naturalness and similarity are shown in Table~\ref{tab:mos}. As can be seen from the N-MOS data of the subjective experiments, it is clear that QuickVC-nosr and QuickVC-sr outperform all baseline models in terms of speech naturalness. In particular, QuickVC-sr achieves the best performance, proving that data augmentation helps the models to better extract content features for speech generation.

In terms of similarity, QuickVC-nosr and QuickVC-sr achieved similar scores to real speech in both S-MOS and Resemblyzer results, with QuickVC-sr scoring higher. Both QuickVC and Diff-VCTK achieved competitive similarity scores, but QuickVC was better in naturalness, and the results of subsequent intelligibility and inference speed experiments showed that QuickVC was much better than Diff-VCTK.

\begin{table}[t]
\centering
\caption{Comparison of naturalness MOS (N-MOS), similarity MOS (S-MOS) with 95\% confidence intervals, and objective speaker similarity score obtained by Resemblyzer.}
\label{tab:mos}
\begin{tabular}{c|c|cc} 
\hline
                                  &N-MOS & S-MOS & Resemblyzer           \\ 
\hline
Diff-VCTK                            & 3.46$\pm$0.21                              & 3.34$\pm$0.21                               & \textbf{87.92\%}               \\
BNE-PPG-VC                            & 3.83$\pm$0.17                              & 2.38$\pm$0.19                               & 78.77\%               \\
VQMIVC                            & 2.68$\pm$0.24                              & 2.07$\pm$0.26                               & 62.04\%               \\ 
\hline
QuickVC-nosr                           & 4.03$\pm$0.15                              & 3.21$\pm$0.17                               & 84.40\%               \\
QuickVC-sr                        & \textbf{4.28$\pm$0.15}                              & \textbf{3.58$\pm$0.20}                               & 85.68\%              \\ 
\hline
Ground Truth     & 4.22$\pm$0.15                              & 3.46$\pm$0.24                              & 81.32\%               \\
\hline
\end{tabular}
\end{table}


\subsubsection{Speech intelligibility}

Word error rate (WER) and character error rate (CER) between source and converted speech represent the intelligibility of the generated speech. It can be seen in Table~\ref{tab:wer} that our proposed models achieve lower WER and CER than all baseline models. This indicates that the proposed method can preserve the linguistic content of source speech well. Besides, we observe that training with data augmentation improves speech intelligence slightly.

\begin{table}[th]
\centering
\caption{Comparison of  intelligibility (WER, CER).}
\label{tab:wer}
\begin{tabular}{c|c|c} 
\hline
           & WER     & CER      \\ 
\hline
Diff-VCTK     & 54.39\% & 30.07\%                                                                \\
BNE-PPG-VC     & 2.81\% & 0.97\%                                                                  \\
VQMIVC     & 66.18\% & 41.01\%  \\
\hline
QuickVC-nosr    & 3.14\% & 1.17\%                                                                 \\
QuickVC-sr & \textbf{2.43\%} & \textbf{0.86\%}                                                              \\
\hline
\end{tabular}
\end{table}

\subsubsection{Speed of voice conversion}
Table~\ref{tab:speed} shows that the inference speed of our approach is the most competitive. Our model is able to generate samples at over 5000 KHz on the 3090 GPU and over 250 KHz on the i9-10900K CPU. Considering that our method is fast while maintaining a high level of naturalness, our methods have an obvious advantage when used in real-time VC systems.

\begin{table}[th]
\centering
\caption{Comparison of  inference speed(in KHz).}
\label{tab:speed}
\begin{tabular}{c|c|c} 
\hline
           & \begin{tabular}[c]{@{}c@{}}{Inference}\\{Speed}\\{GPU}\end{tabular} & \begin{tabular}[c]{@{}c@{}}{Inference}\\{Speed}\\{CPU}\end{tabular}  \\ 
\hline
Diff-VCTK     & 63.34                                                                 & 2.31                                                                  \\
BNE-PPG-VC    & 33.85                                                                 & 22.68                                                                  \\
VQMIVC    & 148.27                                                                 & 123.73                                                                  \\ 
\hline
QuickVC   & \textbf{5320.78}                                                                 & \textbf{280.00}                                                                  \\
\hline
\end{tabular}
\end{table}



%
%

\subsubsection{Latent embedding visualization}

\begin{figure}[th]
    \begin{minipage}[t]{0.5\linewidth}
        \centering
        \includegraphics[width=\textwidth]{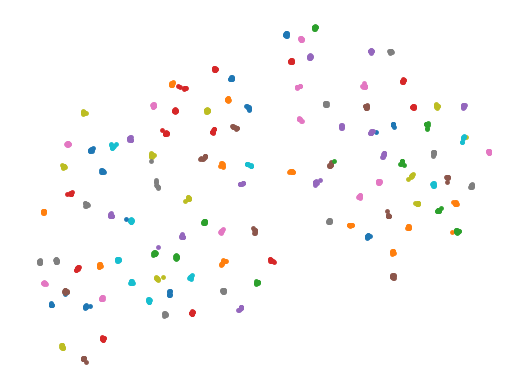}
        \centerline{(a) QuickVC-nosr}
    \end{minipage}%
    \begin{minipage}[t]{0.5\linewidth}
        \centering
        \includegraphics[width=\textwidth]{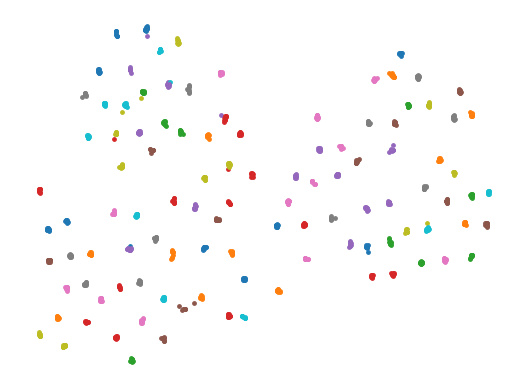}
        \centerline{(b) QuickVC-sr}
    \end{minipage}
    \caption{The Barnes-Hut t-SNE visualization of the speaker embeddings.}
    \label{fig:speaker_embedding}
\end{figure}

The speaker embedding $g$ is the output of the speaker encoder in the QuickVC model. In Figure~\ref{fig:speaker_embedding} we plot the speaker embedding $g$ of 107 speakers in the VCTK dataset using the Barnes-Hut t-SNE method \cite{van2014accelerating}. Each speaker has about 10 audio samples. Based on the speaker embeddings, utterances of the same speaker are clustered together, while those of different speakers are well separated. The 2D speaker embedding demonstrates that the speaker encoder of QuickVC-sr and QuickVC-nosr successfully extracts speaker information.

\section{Conclusion}

In this study, we propose a VITS-based voice conversion model called QuickVC. To achieve fast and high-quality voice conversion, we choose to extract speech content features using HuBERT-Soft and reconstruct the speech using VITS. At the same time, to make the model capable of fast speech synthesis for practical on-device applications, we refer to MS-iSTFT-VITS, reduce the redundancy of decoder computation in the VITS backbone by iSTFT, and adopt a multi-band parallel strategy. The experimental results show a competitive quality of speech synthesized by the model and a competitive level of synthesis speed. However, in the case of zero-shot VC, the model shows a significant decrease in speaker similarity. Future work will mainly focus on this part.



\bibliographystyle{IEEEtran}
\bibliography{mybib}

\end{document}